\documentclass[prl,twocolumn,superscriptaddress,floatfix,shortbibliography]{revtex4-2}
\usepackage{graphicx}
\usepackage{bm}
\usepackage{color,soul}
\usepackage{amsmath,amssymb,amsfonts,float,graphics,epsfig,epstopdf,color,verbatim,tabularx,bm,multirow,appendix,hyperref}

\begin{document}
\title{In-plane anisotropic response to the uniaxial pressure in the hidden order state of URu$_2$Si$_2$}
\author{Xingyu Wang}
\affiliation{Beijing National Laboratory for Condensed Matter Physics, Institute of Physics, Chinese Academy of Sciences, Beijing 100190, China}
\affiliation{School of Physical Sciences, University of Chinese Academy of Sciences, Beijing 100190, China}
\author{Dongliang Gong}
\affiliation{Beijing National Laboratory for Condensed Matter Physics, Institute of Physics, Chinese Academy of Sciences, Beijing 100190, China}
\author{Bo Liu}
\affiliation{Beijing National Laboratory for Condensed Matter Physics, Institute of Physics, Chinese Academy of Sciences, Beijing 100190, China}
\affiliation{School of Physical Sciences, University of Chinese Academy of Sciences, Beijing 100190, China}
\author{Xiaoyan Ma}
\affiliation{Beijing National Laboratory for Condensed Matter Physics, Institute of Physics, Chinese Academy of Sciences, Beijing 100190, China}
\affiliation{School of Physical Sciences, University of Chinese Academy of Sciences, Beijing 100190, China}
\author{Jinyu Zhao}
\affiliation{Beijing National Laboratory for Condensed Matter Physics, Institute of Physics, Chinese Academy of Sciences, Beijing 100190, China}
\affiliation{School of Physical Sciences, University of Chinese Academy of Sciences, Beijing 100190, China}
\author{Pengyu Wang}
\affiliation{Beijing National Laboratory for Condensed Matter Physics, Institute of Physics, Chinese Academy of Sciences, Beijing 100190, China}
\affiliation{School of Physical Sciences, University of Chinese Academy of Sciences, Beijing 100190, China}
\author{Yutao Sheng}
\affiliation{Beijing National Laboratory for Condensed Matter Physics, Institute of Physics, Chinese Academy of Sciences, Beijing 100190, China}
\affiliation{School of Physical Sciences, University of Chinese Academy of Sciences, Beijing 100190, China}
\author{Jing Guo}
\affiliation{Beijing National Laboratory for Condensed Matter Physics, Institute of Physics, Chinese Academy of Sciences, Beijing 100190, China}
\affiliation{Songshan Lake Materials Laboratory , Dongguan, Guangdong 523808, China}
\author{Liling Sun}
\affiliation{Beijing National Laboratory for Condensed Matter Physics, Institute of Physics, Chinese Academy of Sciences, Beijing 100190, China}
\affiliation{School of Physical Sciences, University of Chinese Academy of Sciences, Beijing 100190, China}
\affiliation{Songshan Lake Materials Laboratory , Dongguan, Guangdong 523808, China}
\author{Wen Zhang}
\affiliation{State Key Laboratory of Environment-friendly Energy Materials, Southwest University of Science and Technology, Mianyang, Sichuan 621010, P. R. China}
\affiliation{Science and Technology on Surface Physics and Chemistry Laboratory, Mianyang 621908, China}
\author{Xinchun Lai}
\affiliation{Science and Technology on Surface Physics and Chemistry Laboratory, Mianyang 621908, China}
\author{Shiyong Tan}
\email{tanshiyong@caep.cn}
\affiliation{Science and Technology on Surface Physics and Chemistry Laboratory, Mianyang 621908, China}
\author{Yi-feng Yang}
\email{yifeng@iphy.ac.cn}
\affiliation{Beijing National Laboratory for Condensed Matter Physics, Institute of Physics, Chinese Academy of Sciences, Beijing 100190, China}
\affiliation{School of Physical Sciences, University of Chinese Academy of Sciences, Beijing 100190, China}
\affiliation{Songshan Lake Materials Laboratory , Dongguan, Guangdong 523808, China}
\author{Shiliang Li}
\email{slli@iphy.ac.cn}
\affiliation{Beijing National Laboratory for Condensed Matter Physics, Institute of Physics, Chinese Academy of Sciences, Beijing 100190, China}
\affiliation{School of Physical Sciences, University of Chinese Academy of Sciences, Beijing 100190, China}
\affiliation{Songshan Lake Materials Laboratory , Dongguan, Guangdong 523808, China}
\begin{abstract}
We studied the uniaxial-pressure dependence of the resistivity for URu$_{2-x}$Fe$_x$Si$_2$ samples with $x$ = 0 and 0.2, which host a hidden order (HO) and a large-moment antiferromagnetic (LMAFM) phase, respectively. For both samples, the elastoresistivity $\zeta$ shows a seemingly divergent behavior above the transition temperature $T_0$ and a quick decrease below it. We found that the temperature dependence of $\zeta$ for both samples can be well described by assuming the uniaxial pressure effect on the gap or certain energy scale except for $\zeta_{(110)}$ of the $x$ = 0 sample, which exhibits a non-zero residual value at 0 K. We show that this provides a qualitative difference between the HO and LMAFM phases. Our results suggest that there is an in-plane anisotropic response to the uniaxial pressure that only exists in the hidden order state without necessarily breaking the rotational lattice symmetry.
\end{abstract}


\maketitle

The hidden order (HO) in URu$_2$Si$_2$ has raised great interests in the last three decades \cite{PalstraTTM85,MapleMB86,SchlabitzW86,MydoshJA11,MydoshJA14,MydoshJA20}. The mystery is that a large specific-heat anomaly is observed at the HO transition temperature ($T_{0}$ = 17.5 K), unambiguously suggesting a second-order transition, but no consensus on the order parameter has yet been reached. An antiferromagnetic (AFM) order \cite{BroholmC87} has been found at $T_{0}$, but the ordered moment is too small to account for the large entropy released by the HO phase. Recently, a compelling scenario for the origin of the HO has been proposed based on the in-plane rotational symmetry breaking \cite{OkazakiR11,TonegawaS14,RiggsSC15,KambeS15}. Accordingly, the HO order parameter should spontaneously break the four-fold symmetry and a tetragonal-orthorhombic structural change is expected at $T_0$, which is similar to the nematic order observed in iron-based superconductors \cite{ChuJH12,KuoHH16,HosoiS16,LiuZ16,GuY17}. 
However, more recent studies suggest that there is no evidence for either the structural transition or nematic fluctuations \cite{TabataC14,KambeS18,ChoiJ18,WangL20}.  What seems to be consensus now is that URu$_2$Si$_2$ indeed shows in-plane anisotropic properties under uniaxial strain or magnetic field \cite{OkazakiR11,RiggsSC15,WangL20}. It is thus crucial to ask whether there is some kind of anisotropy that is unique to the HO and if it is, whether it can exist without breaking the rotational symmetry.

In the URu$_2$Si$_2$ system, the so-called large-moment antiferromagnetic (LMAFM) phase has been found to be intimately related to HO. In this phase, the magnetic order has the same ordering vector as the AFM order in HO but with much larger ordered moment. The first-order transition from the HO to the LMAFM phase is first found in URu$_2$Si$_2$ under hydrostatic pressure \cite{AmitsukaH99,JeffriesJR07,JoYJ07,MotoyamaG08,ButchNP10,HassingerE10} and later in Fe-doped samples \cite{KanchanavateeN11,DasP15,HallJS15,WolowiecCT16,RanS16}. Surprisingly, transport and thermodynamic properties have very similar behaviors in both phases although their order parameters are presumably different. Moreover, the Fermi surfaces of these two phases and magnetic excitations are rather similar \cite{HassingerE10,FrantzeskakisE21,WilliamsTJ17}. The crossover between these two phases has been referred to as "adiabatic continuity" since they are intimately related \cite{JoYJ07}.  It is thus intriguing to check the anisotropic behaviors in both the HO and LMAFM phases to find out the unique one of the former, if it does exist.

In this work, we studied the uniaxial elastoresistance $\zeta$ in URu$_{2-x}$Fe$_x$Si$_2$ ($x$ = 0 and 0.2). The divergent behavior of $\zeta$ above $T_0$ was observed in both samples and the amplitude of $\zeta$ shows no significant difference among these samples. We provide a simple explanation based on the uniaxial pressure effect on certain energy scales and thus rule out the existence of nematic fluctuations. Interestingly, a large residual $\zeta$ at 0 K along the (110) direction is only observed in the $x$ = 0 sample, which cannot be described by the above scenario. Our results thus provide a unique property associated with the HO state, which does not necessarily break the lattice rotational symmetry.

Single crystals of URu$_{2-x}$Fe$_x$Si$_2$ were grown by the Czochralski method in a tetra-arc furnace with a continuously purified Ar atmosphere and subsequently annealed at 900 $^{\circ}$C under ultrahigh vacuum for 10 days. The samples were cut into thin plates by a wire saw machine with the desired directions determined by the Laue technique. With the tetragonal notation, the (100) direction corresponds to a or b axis of the lattice while the (110) direction corresponds to the diagonal direction of a and b axes. The elastoresistance was measured by a home-made uniaxial-pressure device based on piezobender as reported previously \cite{LiuZ16,GuY17}.  The uniaxial pressure is defined as positive when the sample is compressed. Since the resistance linearly depends on $p$, we define $\zeta$ = $R(0)^{-1}d\Delta R/dp$, where $R(0)$ is the resistance at zero pressure and $\Delta R$ = $R(p) - R(0)$ \cite{LiuZ16,GuY17}. For all the samples measured here, the maximum uniaxial pressure applied is smaller than 10 MPa. Hydrostatic pressure resistance measurements were performed in a piston/cylinder-type high-pressure cell with Silicone oil as pressure transmitting medium. Pressure is determined by the pressure dependent superconducting transition temperature $T_C$ of Pb \cite{EilingA81} that is placed in the Teflon capsule together with the sample. In the following, we will use $p$ and $P$ to denote the uniaxial pressure and hydrostatic pressure, respectively.

\begin{figure}[tbp]
\includegraphics[width=\columnwidth]{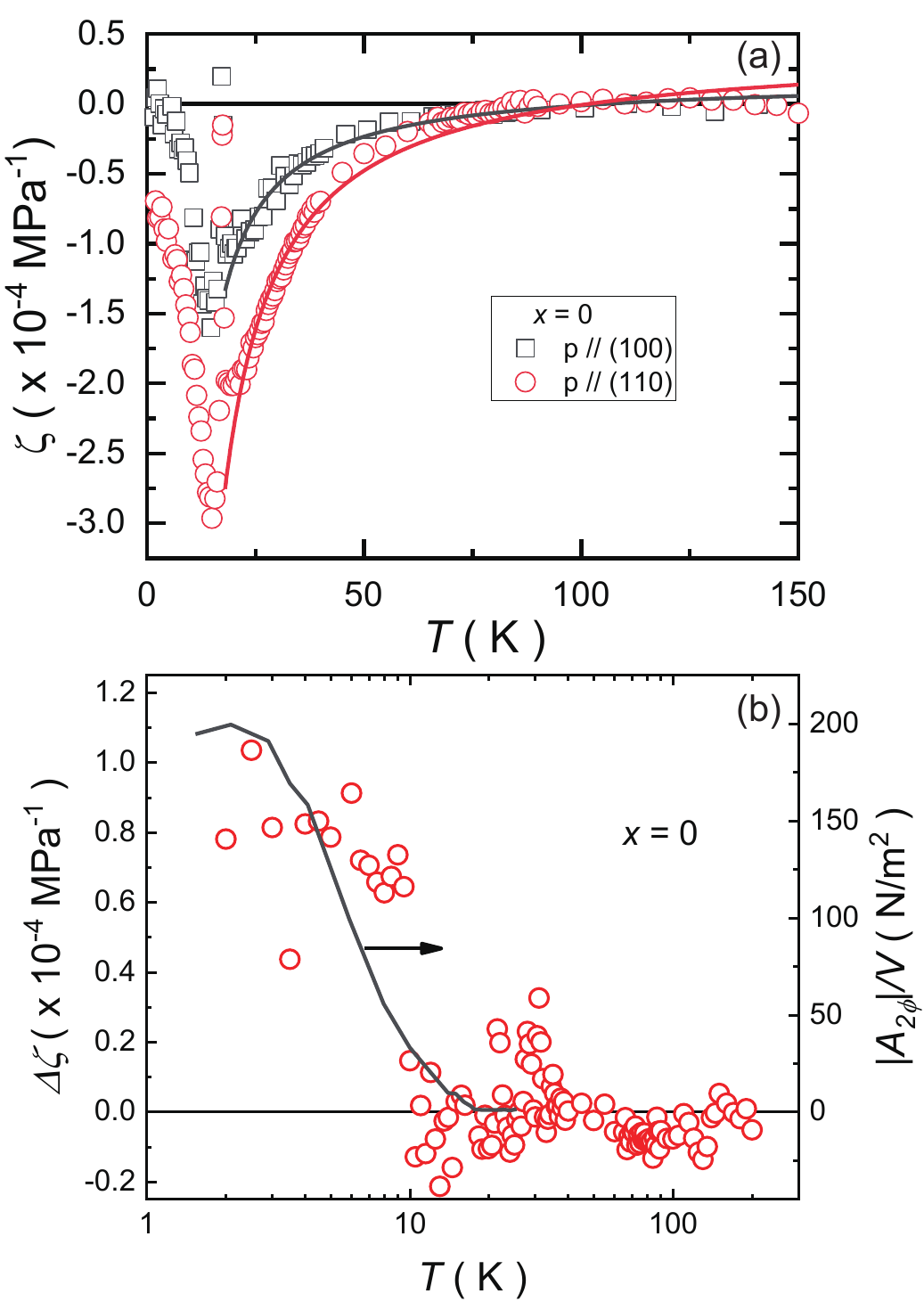}
 \caption{(a) The temperature dependence of $\zeta$ for URu$_2$Si$_2$. The solid lines are fitting results by the Curie-Weiss-like function, as described in the text. (b) Temperature dependence of $\Delta \zeta$ (circles) and $|A_{2\phi}|/V$ (sold line) \cite{OkazakiR11} for URu$_2$Si$_2$. The data around $T_0$ for $\Delta\zeta$ have been omitted for clarification.
}
\label{fig1}
\end{figure}

Figure \ref{fig1}(a) shows the temperature dependence of $\zeta$ for URu$_2$Si$_2$, with the uniaxial pressure $p$ applied along the (100) and (110) directions. For both directions, $\zeta$ exhibits the same temperature dependence. The sharp peak at $T_0$ is associated with the change of $T_0$ under pressure. Similar to previous reports \cite{RiggsSC15}, the data above $T_0$ can be fitted by a Curie-Weiss-like function $A/(T-T')+y_0$, where $A$, $T'$ and $y_0$ are all temperature-independent parameters. The values of $T'$ for both fittings are about 6 K, which is well below $T_0$. It should be pointed out that $T'$ can change significantly for different temperature fitting ranges.

\begin{figure}[tbp]
\includegraphics[width=\columnwidth]{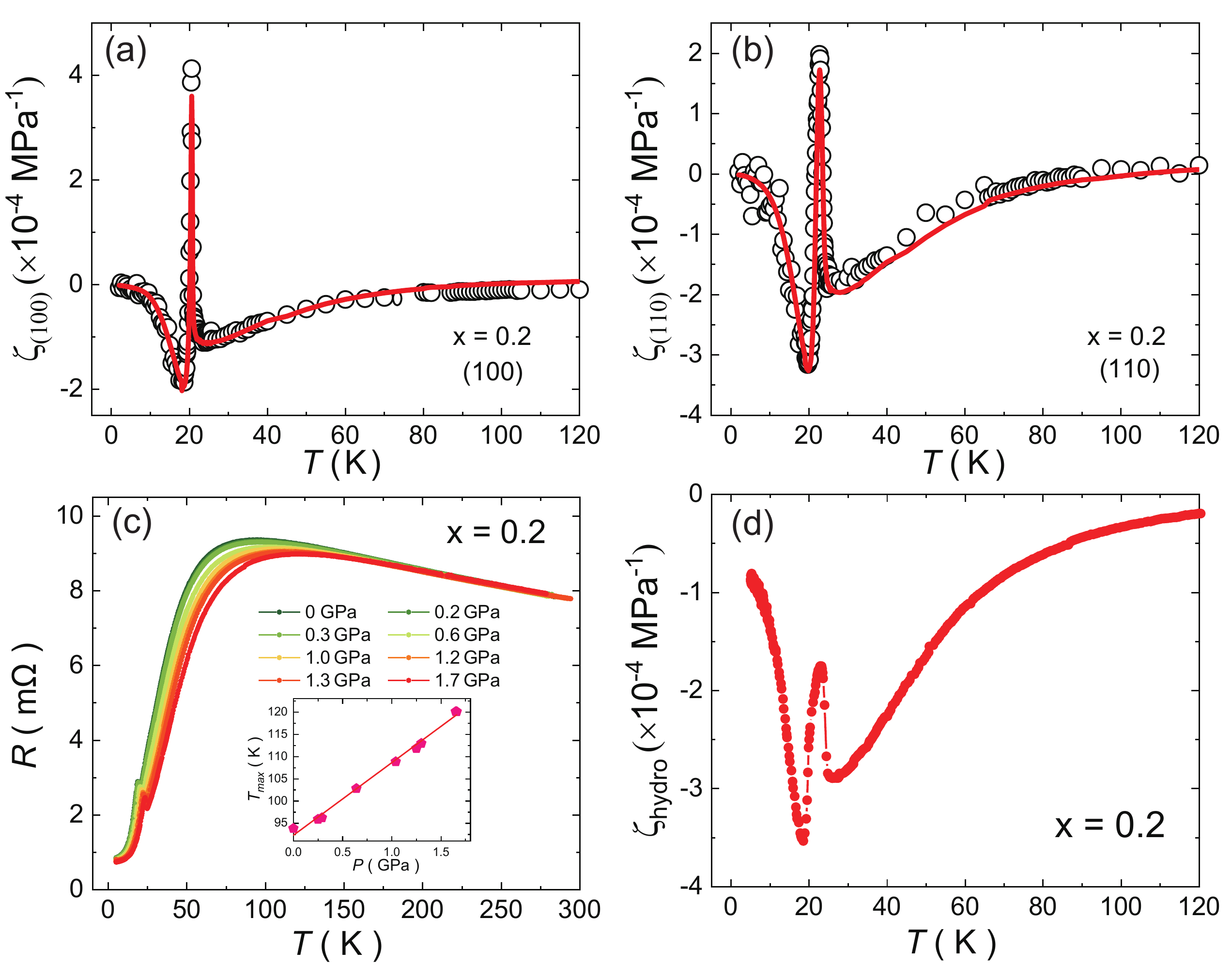}
 \caption{(a) and (b) The temperature dependence of $\zeta_{(100)}$ and $\zeta_{(110)}$, respectively, for the $x$ = 0.2 sample. The solid lines are the calculated results of Eq. (\ref{eqzeta}) with $\delta$ = 6.5 $\times$ 10$^{-5}$ and 1.15 $\times$ 10$^{-4}$ for the (100) and (110) directions, respectively. (c) The temperature dependence of the resistance for the 0.2 sample under different hydrostatic pressures. The inset shows the change of $T_{max}$ under pressure. (d) The temperature dependence of $\zeta_{hydro}$ for the $x$ = 0.2 sample. 
}
\label{fig2}
\end{figure}

A significant difference between $\zeta_{(110)}$ and $\zeta_{(100)}$ is that the former has a finite value at 0 K while the latter tends to go to zero, as shown in Fig. \ref{fig1}(a). This difference has also been observed in the uniaxial strain dependence of the elastoresistivity \cite{RiggsSC15}. Since $\alpha$, the ratio of $\zeta_{(110)}/\zeta_{(100)}$ above $T_0$, is temperature-independent and close to 2 for these two samples, we can distinguish the anisotropy of the elastoresistance in the HO state from that above $T_0$ by calculating $\Delta\zeta$ = $\alpha$$\zeta_{(100)}$ - $\zeta_{(110)}$, as shown in Fig. \ref{fig1}(b). We note that although $\alpha$ may depend on the samples due to their quality and the precision in determining the cross-section area in our method \cite{LiuZ16}, $\Delta\zeta$ above $T_0$ can be reduced to zero with proper choice of $\alpha$. On the other hand, $\Delta\zeta$ always shows a residual value at 0 K no matter what value of $\alpha$ is chosen.  Its temperature dependence seems to be similar to that of the two-fold oscillation amplitude $|A_{2\phi}|/V$ observed in torque measurements \cite{OkazakiR11}.

We will demonstrate that the residual $\Delta\zeta$ at 0 K represents the true anisotropic properties of the HO state. Figures \ref{fig2}(a) and \ref{fig2}(b) show the temperature dependence of $\zeta_{(100)}$ and $\zeta_{(110)}$, respectively, for the $x$ = 0.2 sample. The overall features are similar to those in the $x$ = 0 sample, including the seemingly divergent behavior above $T_0$, the sharp peak at $T_0$ and the quick decrease below $T_0$. However, unlike the $x$ = 0 sample, $\zeta$ tends to go to zero for both directions, which results in a zero $\Delta\zeta$ at 0 K. Since the $x$ = 0.2 sample has the LMAFM ground state \cite{RanS16}, the non-zero $\Delta\zeta$, i.e., the non-zero $\zeta_{(110)}$ at 0 K, should be a unique property of the HO.

Before further discussing the non-zero $\zeta{(110)}$ at 0 K, we address the behaviors of $\zeta$ above $T_0$ first. Previous studies have invoked nematic fluctuations to account for the increase of elastoresistivity with decreasing temperature \cite{RiggsSC15}, which would require the breaking of the in-plane rotational symmetry in the HO state. It was later pointed out that the change of resistivity in URu$_2$Si$_2$ under hydrostatic pressure shows very similar behaviors with that under uniaxial strain, which was attributed to the volume effect \cite{WangL20}. Here we also carried out the resistivity measurements on the $x$ = 0.2 sample under hydrostatic pressure, as shown in Fig. \ref{fig2}(c). Accordingly, we can define $\zeta_{hydro}$ = $R(0)^{-1}\Delta R/\Delta P$, which is shown in Fig. \ref{fig2}(d). Similar to URu$_2$Si$_2$ \cite{WangL20}, the overall features and values are comparable with the uniaxial results, which is against the existence of nematic fluctuations. Even if the elastoresistivity above $T_0$ does come from nematic fluctuations, nematicity should play little role in the formation of the HO since it results in similar behaviors of $\zeta$ above $T_0$ for the LMAFM state.

In fact, we can give a simple explanation for the temperature dependence of $\zeta$ above $T_0$. As seen in Figs. \ref{fig2}(a) and \ref{fig2}(b), $\zeta$ becomes significant only below about 80 K. It is well known that the resistivity across this temperature is associated with the coherence temperature $T^*$ \cite{YangYF08}. Supposing that the resistivity takes the form $\rho=f(T/T^*)$ and the change of $T^*$ by a small pressure $p$ leads to $T^* \rightarrow (1-\delta)T^*$, the change of the resistivity is thus $f[T/(1 - \delta)T^*]-f(T/T^*)$, which is approximately equal to $f[(1 + \delta)T/T^*]-f(T/T^*)$ as $\delta$ is small. Therefore, we can simply change the temperature $T$ to $(1+\delta)T$ in the $R(T)$ data and have the following equation to obtain the elastoresistance,
\begin{equation}
\zeta^* = \frac{R[(1+\delta)T]-R(T)}{R(T)}.
\label{eqzeta}
\end{equation}
\noindent where the unit of $\zeta^*$ is set to MPa$^{-1}$.  It is interesting to note that similar method has been used in removing the phonon contribution of the specific heat if its temperature dependence is only associated with the Debye temperature \cite{BouvierM91,GD16,SongP21}. We also note that Eq. (\ref{eqzeta}) also suggests a close connection between $\zeta$ and $dR/dT$ according to its Taylor expansion \cite{supp}. 

It is clear that this method describes the temperature dependence of $\zeta$ above $T_0$ of the $x$ = 0.2 sample for both the (100) and (110) directions, as shown in Figs. \ref{fig2}(a) and \ref{fig2}(b), respectively. What is rather surprising is that $\zeta$ below $T_0$ and even right at the transition can also be well described by Eq. (\ref{eqzeta}). According to Eq. (\ref{eqzeta}), the value of $\delta$ is associated with the relative change of $T^*$ under the uniaxial pressure. Treating $T^*$ roughly as $T_{max}$, which corresponds to the temperature where the resistance is maximum, the value of $\delta T_{max}/p$, where $p$ is 1 MPa,  should be close to $dT_{max}/dp$. We find that $\delta T_{max}/p$ is about 6.1 and 10.8 K/GPa for the (100) and (110) directions, respectively, which are indeed close to the value of $dT_{max}/dP$ $\sim$ 16 K/GPa under the hydrostatic pressure, as shown in the inset of Fig. \ref{fig2}(c).

\begin{figure}[tbp]
\includegraphics[width=\columnwidth]{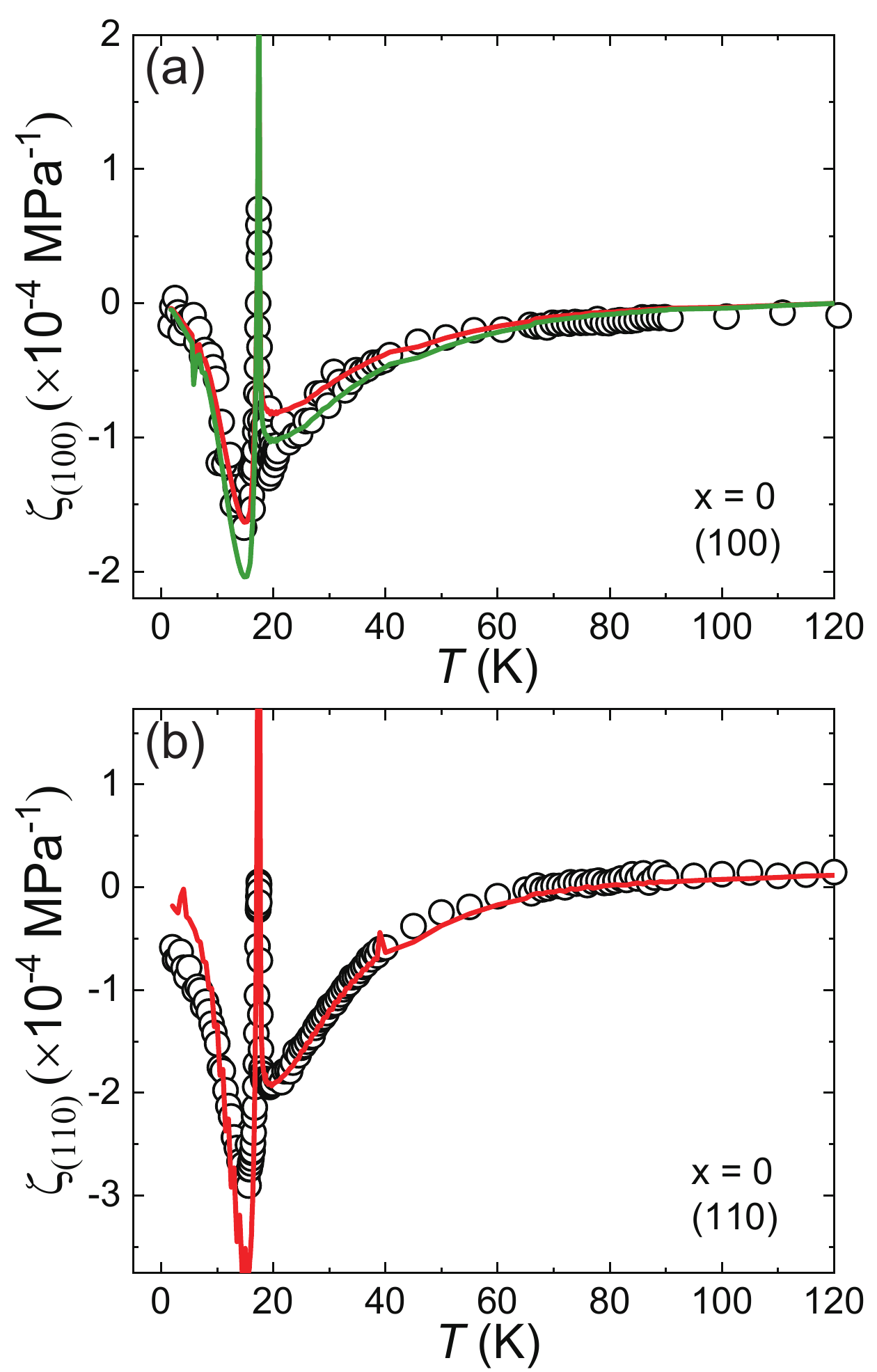}
 \caption{ (a) and (b) The temperature dependence of $\zeta{(100)}$ and $\zeta{(110)}$ for URu$_2$Si$_2$, respectively. The solid lines are the calculated results of Eq. (\ref{eqzeta}) with $\delta$ = 5 $\times$ 10$^{-5}$ and 6 $\times$ 10$^{-5}$ for the red and green lines in (a), respectively, and 1.05 $\times$ 10$^{-4}$ in (b).
}
\label{fig3}
\end{figure}

Now we apply the above analysis to the uniaxial elastoresistivity in the $x$ = 0 sample, as shown in Figs. \ref{fig3}(a) and \ref{fig3}(b). Using Eq. (\ref{eqzeta}), the data above $T_0$ can be again nicely replicated. The value of $\delta T_{max}/p$ is about 4.6 and 8 K/GPa for the (100) and (110) directions, respectively, which are again close to the value of $dT_{max}/dP$ under the hydrostatic pressure ($\sim$ 14 K/GPa) \cite{McElfreshMW87}. Different from the $x$ = 0.2 sample, Eq. (\ref{eqzeta}) cannot describe $\zeta_{(100)}$ for the whole temperature range using just one $\delta$. Rather, we have to change the value of $\delta$ to reproduce $\zeta_{(100)}$ below $T_0$, which seems to be associated with the change of the gap in the HO state under pressure \cite{McElfreshM87,RanS16,supp}. However, we can never describe $\zeta_{(110)}$ below $T_0$ by Eq. (\ref{eqzeta}) due to the existence of the residual 0-K value. This again demonstrates the unique connection between $\Delta\zeta$ and the HO. 

Our results show that the large non-zero value of $\zeta_{(110)}$ at zero K represents the intrinsic anisotropic response of the resistivity to the uniaxial pressure in the HO state. We note that this behavior is different from that of the uniaxial-pressure-induced AFM moment, which shows no difference between the (100) and (110) directions \cite{YokoyamaM02}. It is commonly believed that the resistivity below $T_0$ can be described by gap functions for both the HO and LMAFM phases \cite{McElfreshM87,RanS16}. This is consistent with Eq. (\ref{eqzeta}) as the gap $\Delta$ in the functions is always coupled to the temperature as $\Delta/T$ \cite{supp}. Our results of $\zeta_{(110)}$ for the $x$ = 0 sample suggest that there should exist a component of the resistivity, which can be only invoked under uniaxial pressure along the (110) direction, that is beyond any kind of gap functions \cite{McElfreshM87,RanS16}. This suggests that the unique information of the HO state may be only revealed under uniaxial pressure. 

Previous studies have shown some quantitative differences between the HO and LMAFM states \cite{JoYJ07,MotoyamaG08,WolowiecCT16,RanS16}. Our results show that the HO state can be {\it qualitatively} distinguished from the LMAFM state by their response to the uniaxial pressure along the (110) direction. 
It is particularly needed to emphasize that contrary to some previous results \cite{OkazakiR11,TonegawaS14,RiggsSC15,KambeS15}, the above uniaxial-pressure-induced anisotropy does not require breaking the rotational symmetry \cite{TabataC14,KambeS18,ChoiJ18,WangL20}. We note that in the case when the sample is not free-standing, the local residual stress induced by glue, etc., may effectively provide uniaxial pressure or strain to give rise to results that should be absent at zero strain \cite{RenX15}. This could explain the similar behaviors between $\Delta\zeta$ and $|A_{2\phi}|/V$ in Fig. \ref{fig1}(b).  Our results may thus suggest a somehow strange property of the HO. On the one hand the HO does not break the rotational symmetry, so our results are not conducive to those theoretical proposals that require the rotational symmetry breaking \cite{HauleK09,CricchioF09,IkedaH12,RauJG12,HsuCH14}. But on the other hand the proposed HO order parameter should be coupled to the $B_{2g}$ channel due to the symmetry constraint revealed by our measurements. It may be a requisition for a theory to study the proposed order parameter in the presence of a slight lattice distortion along the (110) direction.

In conclusion, we show that there are no nematic fluctuations in URu$_2$Si$_2$ and a large residual $\zeta_{(110)}$ exists at 0 K. The latter is the unique feature to the HO since it is not found in the LMAFM state. Our results suggest that the HO can exhibit large anisotropic response to the uniaxial pressure or strain without rotational symmetry breaking, which could provide an interesting perspective in searching the order parameter of the HO.

\acknowledgments

This work is supported by the National Key Research and Development Program of China (Grants No. 2017YFA0302903, No. 2017YFA0303100, No. 2020YFA0406003, No. 2021YFA1400401), the National Natural Science Foundation of China (Grants No. 11961160699, No. 11874401,11974397), the Strategic Priority Research Program(B) of the Chinese Academy of Sciences (Grants No. XDB33000000, No. XDB25000000), the K. C. Wong Education Foundation (GJTD-2020-01).

\end{document}